\documentclass[english,aps,preprint]{revtex4}
\usepackage[T1]{fontenc}
\usepackage[latin9]{inputenc}
\usepackage{amsmath}
\usepackage{amssymb}
\usepackage{esint}

\makeatletter
\@ifundefined{textcolor}{}
{%
 \definecolor{BLACK}{gray}{0}
 \definecolor{WHITE}{gray}{1}
 \definecolor{RED}{rgb}{1,0,0}
 \definecolor{GREEN}{rgb}{0,1,0}
 \definecolor{BLUE}{rgb}{0,0,1}
 \definecolor{CYAN}{cmyk}{1,0,0,0}
 \definecolor{MAGENTA}{cmyk}{0,1,0,0}
 \definecolor{YELLOW}{cmyk}{0,0,1,0}
}

\usepackage{babel}

\makeatother

\usepackage{babel}
\begin{document}

\title{Some Generalizations in Supersymmetric Quantum Mechanics and the
Supersymmetric $\varepsilon$-System Revisited}

\author{E. A. Gallegos}

\email{egallegoscollado@gmail.com}

\selectlanguage{english}%

\affiliation{Departamento de Física, Universidade Federal de Santa Catarina, Campus
Trindade, 88040-900, Florianópolis, SC, Brazil}

\author{A. J. da Silva}

\email{ajsilva@fma.if.usp.br}

\selectlanguage{english}%

\affiliation{Instituto de Física, Universidade de São Paulo, Caixa Postal 66318,
05315-970, São Paulo, SP, Brazil }

\author{D. Spehler}

\email{spehler@iphc.cnrs.fr}

\selectlanguage{english}%

\affiliation{IPHC-DRS, UdS, CNRS-in2p3, 23 Rue du Loess, 67 037, Strasbourg, France}
\begin{abstract}
We discuss two distinct aspects in supersymmetric quantum mechanics.
First, we introduce a new class of operators $A$ and $\bar{A}$ in
terms of anticommutators between the momentum operator and $N+1$
arbitrary superpotentials. We show that these operators reduce to
the conventional ones which are the starting point in standard supersymmetric
quantum mechanics. In this context, we argue furthermore that supersymmetry
does not only connect Schrödinger-like operators, but also a more
general class of differential operators. Second, we revisit the supersymmetric
$\varepsilon$-system recently introduced in the literature by exploiting
its intrinsic supersymmetry. Specifically, combining the Hamilton
hierarchy method and the $\delta$-expansion method, we determine
an energy for the first excited state of the bosonic Hamiltonian close
to that calculated in earlier works.
\end{abstract}
\maketitle

\section{Introduction}

Since its invention by Witten \cite{Witten (1981)} in connection
with supersymmetry (SUSY) breaking issues in quantum field theory,
supersymmetric quantum mechanics (SQM) has become an independent and
a fruitful realm of research. Not only old problems such as the hydrogen
atom were rephrased in terms of SUSY \cite{Kostelecky-etal (1985)},
but also more recently its conformal extensions (i.e., superconformal
quantum mechanics) have shown to be relevant in quantum black holes
and AdS/CFT correspondence (see e.g. \cite{AdS-CFT} and references
herein).

As is well known, a supersymmetric system in quantum mechanics is
described by a Hamiltonian which is expressed in terms of a set of
supercharges (at least two). The superalgebra of these elements (Hamiltonian
and supercharges, see Eqs. (\ref{eq:2.1})-(\ref{eq:2.2})) is realized
by introducing two operators $A$ and $\bar{A}$ which in the standard
approach depend merely on the momentum $p$ operator and on a function
$W\left(x\right)$, called superpotential. It is possible to show
that the Hamiltonians $H_{-}$ and $H_{+}$, diagonal entries of the
supersymmetric Hamiltonian, constructed from these operators are Schrödinger-like
operators, being both connected by supersymmetry. A footprint of this
relationship is the isospectral energy structure which both Hamiltonians
possess.

We discuss some generalizations of the standard supersymmetric quantum
mechanics. By introducing a new class of operators $A$ and $\bar{A}$
which reduce to the old ones, we show that supersymmetry does not
only link Schrödinger-like operators, but also a more general class
of differential operators. The $N=1$ case corresponds to supersymmetric
systems described by Shrödinger-like operators and deformations of
them, whereas the $N>1$ case corresponds to supersymmetric systems
with higher-derivative operators. A particular study of this last
case was carried out long ago in \cite{Andrianov-Ioffe (1993)}, concerning
the Witten index (a topological quantity which indicates whether or
not supersymmetry is broken in usual theories).

In this work we also treat the supersymmetric $\varepsilon$-system
recently introduced and studied (using the variational method) in
\cite{Fabricio-Adilson (2012)}. Here we combine the Hamiltonian hierarchy
method \cite{Sukumar (1985)} and the $\delta$-expansion method \cite{Bender (1989),Cooper & Roy (1990)},
for solving the Riccati equations, in order to find the first excited
state energy of the bosonic Hamiltonian. Our result is close to that
found in \cite{Fabricio-Adilson (2012)} and more recently in \cite{Fernandez (2013)}.

The paper is organized as follows. In Sec. \ref{sec: SQM} we review
the ``standard'' supersymmetric quantum mechanics (SQM) and discuss
some generalizations of it. In addition, we treat briefly the Hamiltonian
hierarchy method which will be used in the next section. In Sec. \ref{sec:First-Excited-Estate}
we revisit the supersymmetric $\varepsilon$-system following another
approach. As mentioned above, here we exploit its intrinsic supersymmetry
to face the same energy eigenvalue problem than in \cite{Fabricio-Adilson (2012)}.
Finally, Sec. \ref{sec:Conclunsions} contains our main results.

\section{SUSY Quantum Mechanics And Some Generalizations\label{sec: SQM}}

In this section we review the core of standard supersymmetric quantum
mechanics (SQM) and discuss some generalizations of it. To this end,
we follow closely \cite{Sukumar (1985),Shifman (1995)}. 

The simplest SQM is described in terms of two supercharges $Q$ and
$\bar{Q}$ (its Hermitian adjoint), which obey the following algebra
\cite{Witten (1981)}
\begin{equation}
H=\left\{ Q,\,\bar{Q}\right\} ,\qquad Q^{2}=0=\bar{Q}^{2},\label{eq:2.1}
\end{equation}
where $H$ denotes the Hamiltonian of the supersymmetric system. It
is easy to show, using the above algebra, that the supercharges $Q$,
$\bar{Q}$ are constants of motion, that is
\begin{equation}
\left[Q,\, H\right]=0,\qquad\left[\bar{Q},\, H\right]=0.\label{eq:2.2}
\end{equation}
A simple realization of the algebra (\ref{eq:2.1}) is achieved by
choosing
\begin{equation}
Q=\frac{1}{2}\left(\sigma^{1}-i\sigma^{2}\right)A\qquad\mbox{and}\qquad\bar{Q}=\frac{1}{2}\left(\sigma^{1}+i\sigma^{2}\right)\bar{A},\label{eq:2.3}
\end{equation}
where $\sigma^{1}$ and $\sigma^{2}$ are the usual Pauli matrices
and where $A$ is an arbitrary differential operator ($\bar{A}$ being
its Hermitian adjoint). The supersymmetric Hamiltonian $H$ in (\ref{eq:2.1})
takes the form
\begin{equation}
H=\frac{1}{2}\left(\sigma^{0}+\sigma^{3}\right)\bar{A}A+\frac{1}{2}\left(\sigma^{0}-\sigma^{3}\right)A\bar{A}=\left(\begin{array}{cc}
\bar{A}A & 0\\
0 & A\bar{A}
\end{array}\right),\label{eq:2.4}
\end{equation}
a structure (diagonal) which allows us to identify two distinct but
intimately connected by supersymmetry sectors in the state space of
the system described by $H$. Adopting the notation $H_{-}=\bar{A}A$
and $H_{+}=A\bar{A}$ for the diagonal elements of $H$ and writing
the state function of the system as
\begin{equation}
\Psi\left(x\right)=\left(\begin{array}{c}
\psi^{-}\\
\psi^{+}
\end{array}\right)=\left(\begin{array}{c}
\psi^{-}\\
0
\end{array}\right)+\left(\begin{array}{c}
0\\
\psi^{+}
\end{array}\right),\label{eq:2.5}
\end{equation}
it is straightforward to observe that the two functions on the right
hand side of (\ref{eq:2.5} ) are independent one to another, and
so belong to different sectors of the state space. Notice also that
since $H$ has a diagonal structure, this operator does not ``smear''
the pureness of these kinds of functions. In other words, the operators
$H_{\mp}$ do act on the respective component functions $\psi^{\mp}$
of $\Psi$: 
\begin{equation}
H\Psi=\left(\begin{array}{cc}
H_{-} & 0\\
0 & H_{+}
\end{array}\right)\left(\begin{array}{c}
\psi^{-}\\
\psi^{+}
\end{array}\right)=\left(\begin{array}{c}
H_{-}\psi^{-}\\
0
\end{array}\right)+\left(\begin{array}{c}
0\\
H_{+}\psi^{+}
\end{array}\right).
\end{equation}
In what follows a pure state $\Psi^{-}=\frac{1}{2}\left(\sigma^{0}+\sigma^{3}\right)\Psi$
will be called bosonic and a pure state $\Psi^{+}=\frac{1}{2}\left(\sigma^{0}-\sigma^{3}\right)\Psi$
fermionic. We stress however that ``bosonic'' and ``fermionic''
are simply labels and have nothing to do with the geometrical concept
of spin which does not exist in one-dimensional space.

Before proceeding with the construction of operators $A$ and $\bar{A}$,
there are two direct consequences of the graded algebra (\ref{eq:2.1},
\ref{eq:2.2}) which are worthwhile to mention. The positivity of
energy of a supersymmetric system on the one hand, and the intertwining
relationship between its bosonic and fermionic sectors by means of
the supercharges on the other. The positive feature of the spectrum
becomes evident if we compute, with the aid of the $H$-$Q$ anticommutator
in (\ref{eq:2.1}), the expectation value of $H$ corresponding to
an arbitrary state $\left|\Psi\right\rangle $,
\begin{equation}
E_{\Psi}=\left\langle \Psi\left|H\right|\Psi\right\rangle =\left|Q\left|\Psi\right\rangle \right|^{2}+\left|\bar{Q}\left|\Psi\right\rangle \right|^{2}\geq0,\label{eq:2.6}
\end{equation}
 while the bosonic-fermionic relationship is ascertained by regarding
the $H$-$Q$ commutators (\ref{eq:2.2}) and the own structure of
the supercharges. For example, let $\Psi^{+}=\frac{1}{2}\left(\sigma^{0}-\sigma^{3}\right)\Psi$
be a given fermionic eigenfunction of $H$ with eigenvalue $E_{+}$,
i. e. $H\Psi^{+}=E_{+}\Psi^{+}$. Applying the supercharge $\bar{Q}$
on both sides of this eigenvalue equation, one obtains by using (\ref{eq:2.2})
\begin{equation}
\bar{Q}\left(H\Psi^{+}\right)=H\left(\bar{Q}\Psi^{+}\right)\Rightarrow H\left(\bar{Q}\Psi^{+}\right)=E\left(\bar{Q}\Psi^{+}\right),
\end{equation}
which indicates that the function $\bar{Q}\Psi^{+}$ is also an eigenfunction
of $H$ with the same eigenvalue $E_{+}$ as $\Psi^{+}$. Moreover,
taking into account the peculiar form of $\bar{Q}$ given in (\ref{eq:2.3}),
one easily observes that the eigenfunction $\bar{Q}\Psi^{+}$ has
the structure of a bosonic function $\Psi^{-}$. In fact,
\begin{equation}
\bar{Q}\Psi^{+}=\left(\begin{array}{cc}
0 & \bar{A}\\
0 & 0
\end{array}\right)\left(\begin{array}{c}
0\\
\psi^{+}
\end{array}\right)=\left(\begin{array}{c}
\bar{A}\psi^{+}\\
0
\end{array}\right)\sim\Psi^{-}.
\end{equation}
Hence if $\Psi^{+}$ is a normalized fermionic eigenfunction of $H$
with eigenvalue $E_{+}$, then its normalized bosonic partner $\Psi^{-}$
(also eigenfunction of $H$, with the same eigenvalue as $\Psi^{+}$)
is given by $\Psi^{-}=\left(E_{+}\right)^{-1/2}\bar{Q}\Psi^{+}$.
A similar analysis can be done for $\Psi^{-}$.

So far we have simply provided an overview of standard supersymmetric
quantum mechanics, without nothing new. In what follows let us consider
a variant of this by introducing a general class of operators $A$
and $\bar{A}$. This is possible because of the weak constraints imposed
on such operators, namely, that one must be the Hermitian adjoint
of the other $\left(\bar{\bar{A}}=A\right)$ and that $H_{-}=\bar{A}A$
and $H_{+}=A\bar{A}$ must be positive semi-definite operators.

Since in the $x$-representation $A$ and $\bar{A}$ have to depend
on the basic $x$ and $p=-id/dx$ operators, we propose a finite series
in $p$ with $x$-dependent coefficients for each of them,
\begin{equation}
A(x,\, p)=\frac{1}{2}\sum_{n=0}^{N}a_{n}\left(x\right)\star p^{n}\qquad\qquad\bar{A}(x,\, p)=\frac{1}{2}\sum_{n=0}^{N}\overline{a_{n}\left(x\right)\star p^{n}},\label{eq:2.10}
\end{equation}
where $X\star Y$ denotes the anticommutator of $X$ and $Y$, i.
e., $X\star Y\doteq\left\{ X,\, Y\right\} $. Here $a_{n}\left(x\right)$
are in general $(N+1)$ complex functions of $x$, which we call from
now on generalized superpotentials. In this way, one needs to fix
($N+1$) complex superpotentials in order to specify completely $A$
and its adjoint $\bar{A}$. Note that for consistency the $N=0$ case
must be ruled out. The $N=1$ and $N>1$ cases contain ``dynamic''
in their structures and are rich in possibilities (possible choices
for the functions $a_{n}\left(x\right)$). 

We analyze now the simplest cases, i. e., $N=1$ and $N=2$. We show
below that $N=1$ leads to theories governed by Schrodinger-like equations
and deformations of them, while $N=2$ leads to theories with higher
derivatives.

Taking $N=1$ in (\ref{eq:2.10}), one obtains
\begin{equation}
A=\frac{1}{2}\left(a_{0}\star p^{0}+a_{1}\star p^{1}\right)=a_{0}+\frac{1}{2}\left(pa_{1}\right)+a_{1}p\label{eq:2.11a}
\end{equation}
\begin{equation}
\bar{A}=\frac{1}{2}\left(\overline{a_{0}\star p^{0}}+\overline{a_{1}\star p^{1}}\right)=a_{0}^{*}+\frac{1}{2}\left(pa_{1}^{*}\right)+a_{1}^{*}p.\label{eq:2.11b}
\end{equation}
Here we have made use of the (anti)commutator relations,
\begin{equation}
\overline{\left[X,\, Y\right]}=-\left[\bar{X},\,\bar{Y}\right]\qquad\qquad\overline{\left\{ X,\, Y\right\} }=\left\{ \bar{X},\,\bar{Y}\right\} ,\label{eq:2.13}
\end{equation}
along with the Hermiticity of the momentum operator: $\bar{p}=p=-id/dx$.
Note that the use of the anticommutator property allows us to obtain
from $A$ an adjoint $\bar{A}$ completely symmetric. This is the
reason for which anticommutators were introduced in the definitions
(\ref{eq:2.10}).

Using the explicit form of $A$ and $\bar{A}$ in (\ref{eq:2.11a},
\ref{eq:2.11b}), one sets up easily the $H_{\mp}$ ``Hamiltonians'':
\begin{eqnarray}
H_{-} & = & \left|a_{1}\right|^{2}p^{2}+\left(a_{0}^{*}a_{1}+a_{1}^{*}a_{0}+\frac{3}{2}a_{1}^{*}\left(pa_{1}\right)+\frac{1}{2}a_{1}\left(pa_{1}^{*}\right)\right)p\nonumber \\
 &  & +\left(\left|a_{0}\right|^{2}+\frac{1}{2}a_{0}^{*}\left(pa_{1}\right)+\frac{1}{2}a_{0}\left(pa_{1}^{*}\right)+\frac{1}{4}\left(pa_{1}^{*}\right)\left(pa_{1}\right)+a_{1}^{*}\left(pa_{0}\right)+\frac{1}{2}a_{1}^{*}\left(p^{2}a_{1}\right)\right),\label{eq:2.14a}\\
\nonumber \\
H_{+} & = & \left|a_{1}\right|^{2}p^{2}+\left(a_{0}a_{1}^{*}+a_{1}a_{0}^{*}+\frac{3}{2}a_{1}\left(pa_{1}^{*}\right)+\frac{1}{2}a_{1}^{*}\left(pa_{1}\right)\right)p\nonumber \\
 &  & +\left(\left|a_{0}\right|^{2}+\frac{1}{2}a_{0}^{*}\left(pa_{1}\right)+\frac{1}{2}a_{0}\left(pa_{1}^{*}\right)+\frac{1}{4}\left(pa_{1}^{*}\right)\left(pa_{1}\right)+a_{1}\left(pa_{0}^{*}\right)+\frac{1}{2}a_{1}\left(p^{2}a_{1}^{*}\right)\right).\label{eq:2.14b}
\end{eqnarray}
These second-order linear differential operators become Schrodinger-like
operators only if one chooses adequately the functions $a_{0}(x)$
and $a_{1}(x)$. The right choice at first sight is $a_{0}=a_{0}^{*}$
(real function) and $a_{1}=i/\sqrt{2m}$, since in this way the coefficient
of $p^{2}$ turns out $1/(2m)$ and the $p$-linear terms vanish.
Setting $2m\doteq1$ and $a_{0}(x)\doteq W(x)$, we can write $H_{\pm}$
as
\begin{equation}
H_{\pm}=p^{2}+V_{\pm},\label{eq:2.15a}
\end{equation}
where
\begin{equation}
V_{\pm}=W{}^{2}\pm dW/dx\label{eq:2.15b}
\end{equation}
are known as Riccati's equations. Here the operators $A$ and $\bar{A}$
in (\ref{eq:2.11a},\ref{eq:2.11b}) become simple functions of $W$:
\begin{equation}
A=W\left(x\right)+i\, p\qquad\qquad\bar{A}=W\left(x\right)-i\, p.\label{eq:2.15c}
\end{equation}
These kinds of operators were considered long ago in \cite{Bernstein-etal (1984)}
and actually constitute the starting point of ``standard'' supersymmetric
quantum mechanics. Nevertheless, the liberty of choosing the functions
$a_{0,1}\left(x\right)$ opens the door to regard some modifications
(deformations) of the connected Schrodinger-like equations in SQM.
For instance, taking $a_{0}=a_{0}^{*}=W(x)$ and $a_{1}=i\alpha\left(x\right)$,
with $\alpha\left(x\right)$ real, in (\ref{eq:2.14a},\ref{eq:2.14b}),
one modifies the standard Schrödinger-like Hamiltonians $H_{\pm}$
given in (\ref{eq:2.15a}). In this case, the modified Hamiltonians
$\tilde{H}_{\pm}$ may be written as
\begin{equation}
\tilde{H}_{\pm}=H_{\pm}+\left(\alpha^{2}-1\right)p^{2}+\left[\pm i\left(\alpha-1\right)\left(pW\right)+\frac{1}{4}\left(p\alpha\right)^{2}+\frac{1}{2}\alpha\left(p^{2}\alpha\right)+2\alpha\left(p\alpha\right)p\right],
\end{equation}
where the last two terms modify the kinetic and potential parts of
$H_{\pm}$. The key point in this analysis is that by construction
the operators $\tilde{H}_{-}$ and $\tilde{H}_{+}$ must also be linked
by supersymmetry.

Now we are going to consider briefly the $N=2$ case. From (\ref{eq:2.10}),
setting $N=2$, one can easily verify that
\begin{eqnarray}
A & = & \frac{1}{2}\left(a_{0}\star p^{0}+a_{1}\star p^{1}+a_{2}\star p^{2}\right)\nonumber \\
 & = & a_{2}p^{2}+\left(a_{1}+\left(pa_{2}\right)\right)p+\left(a_{0}+\frac{1}{2}\left(pa_{1}\right)+\frac{1}{2}\left(p^{2}a_{2}\right)\right)\label{eq:2.16a}\\
\nonumber \\
\bar{A} & = & \frac{1}{2}\left(\overline{a_{0}\star p^{0}}+\overline{a_{1}\star p^{1}}+\overline{a_{2}\star p^{2}}\right)\nonumber \\
 & = & a_{2}^{*}p^{2}+\left(a_{1}^{*}+\left(pa_{2}^{*}\right)\right)p+\left(a_{0}^{*}+\frac{1}{2}\left(pa_{1}^{*}\right)+\frac{1}{2}\left(p^{2}a_{2}^{*}\right)\right).\label{eq:2.16b}
\end{eqnarray}
 As these operators are second-order differential ones, $H_{-}=\bar{A}A$
and $H_{+}=A\bar{A}$ will be in general fourth-order differential
operators. Even though $H_{-}$ and $H_{+}$ must still be intimately
connected by supersymmetry, they turn out rather intricate without
the imposition of additional conditions on the superpotentials $a_{i}\left(x\right)$.
A particular case of (\ref{eq:2.16a}, \ref{eq:2.16b}), which is
obtained by putting $a_{0}=a_{0}^{*}=\varphi\left(x\right)$, $a_{1}\left(x\right)=i\, f\left(x\right)$,
and $a_{2}=1$, was studied long ago in \cite{Andrianov-Ioffe (1993)}.
However, as can be easily seen, there are an infinity of possibilities
which can be of interest from the physical or mathematical point of
view.

Coming back to the $N=1$ case and focusing, in particular, on formulas
(\ref{eq:2.15a}-\ref{eq:2.15c}), we show that it is always possible
to write a one dimensional Hamiltonian $H_{-}=p^{2}+V\left(x\right)$
in the form $H_{-}=\bar{A}A+c$, where the operators $A$ and $\bar{A}$
are given in (\ref{eq:2.15c}) and $c$ is an arbitrary constant.
Here we follow the same line of reasoning as in \cite{Sukumar (1985)}.
Writing $H_{-}=\bar{A}A+c$ in terms of the superpotential $W$ with
the help of (\ref{eq:2.15c}) and comparing the result with the standard
form $H_{-}=p^{2}+V\left(x\right)$, one arrives at
\begin{equation}
W^{2}-dW/dx=V\left(x\right)-c.\label{eq:2.17a}
\end{equation}
So the superpotential $W$ which defines the operators $A$ and $\bar{A}$
must be a solution of the above equation. Obviously, this solution
will depend on the form of the energy potential $V\left(x\right)$
and the value of the constant $c$. If now we fix the arbitrariness
of $c$ by equaling it to a given eigenvalue $E$ of $H$, $H\psi_{E}=E\psi_{E}$,
one finds a solution $W_{E}$ of (\ref{eq:2.17a}):
\begin{equation}
W_{E}\left(x\right)=-\frac{1}{\psi_{E}}\frac{d\psi_{E}}{dx}.\label{eq:2.17b}
\end{equation}
Notice that $W_{E}$ is implicitly a function of the eigenvalue $E$
by means of its corresponding eigenfunction $\psi_{E}$. Solving this
differential equation one obtains a way of expressing the eigenfunction
$\psi_{E}$ in terms of its superpotential $W_{E}$:
\begin{equation}
\psi_{E}\left(x\right)=\psi_{E}\left(0\right)e^{-\int_{0}^{x}W_{E}\left(y\right)dy}.\label{eq:2.17c}
\end{equation}
Some comments concerning equations (\ref{eq:2.17a}-\ref{eq:2.17c})
are in order. First, note that (\ref{eq:2.17c}) is true for any given
eigenfunction $\psi_{E}$ with eigenvalue $E$ of $H_{-}=p^{2}+V\left(x\right)$.
However the simplest factorization of the Hamiltonian $H_{-}$, i.e.
$H_{-}=\bar{A}A$, is achieved if and only if one chooses the ground
state $\psi_{E_{0}}$ corresponding to $E_{0}=0$. This is always
possible since in quantum mechanics one can fix the ground state energy
$E_{0}$ to zero by subtracting $E_{0}$ from the Hamiltonian $H$.
Second, the bosonic Hamiltonian $H_{-}=\bar{A}A+c$ has a partner
$H_{+}=A\bar{A}+c$ in such way that they are linked by the supercharges:
$H_{-}\overset{Q\left(\bar{Q}\right)}{\longleftrightarrow}H_{+}$.
Finally, from (\ref{eq:2.17c}) one realizes that the normalizability
of the eigenfunction $\psi_{E}\left(x\right)$ depends on the behavior
of the superpotential $W_{E}\left(x\right)$ when $x\rightarrow\pm\infty$.

From the above analysis, it is evident that one can always associate
a set of Hamiltonians (constructed successively by following the procedure
described in the previous paragraph) to a given Hamiltonian so that
the eigenvalues and eigenfunctions of any two adjacent Hamiltonians
are connected by supersymmetry. This hierarchy of Hamiltonians was
studied for the first time in \cite{Sukumar (1985)} and, as we shall
see in the next section, becomes to be a power tool (in conjunction
with the $\delta$-expansion \cite{Bender (1989),Cooper & Roy (1990)})
in finding approximate eigenvalues and eigenfunctions of a given Hamiltonian.

\section{The Supersymmetric $\varepsilon$-System Revisited\label{sec:First-Excited-Estate}}

In this section we study the supersymmetric $\varepsilon$-system
(of order two in $x$) defined by the superpotential $W\left(x\right)=gx^{2}\varepsilon\left(x\right)$,
with $g>0$. This type of model was recently introduced and studied
in detail in \cite{Fabricio-Adilson (2012)}. Indeed, by means of
the variational technique, the authors in \cite{Fabricio-Adilson (2012)}
computed the approximate energy eigenvalues of the first excited states
of the partner Hamiltonians, establishing explicitly the SUSY relationship
between them. In what follows we revisit the $\varepsilon$-system
and exploit its supersymmetry in order to tackle the same energy eigenvalue
problem. According to the method described in \cite{Cooper & Roy (1990)},
we first modify the superpotential $W$ by introducing an extra $\delta$
parameter in terms of which we will carry out the perturbation expansion:
\begin{equation}
W\left(x\right)=g\left|x\right|^{1+\delta}\varepsilon\left(x\right),\label{eq:3.1}
\end{equation}
where our original superpotential is obviously recovered by taking
$\delta=1$. Notice also that the absolute value of $x$ it is necessary
to guarantee the right behavior of $W(x)$ at infinity: negative sign
of $W\left(x\right)$ at minus infinity and positive sign at plus
infinity.

Inserting (\ref{eq:3.1}) into (\ref{eq:2.15b}), one obtains the
partner potentials
\begin{equation}
V_{\pm}=g^{2}x^{2\left(1+\delta\right)}\pm g\left(1+\delta\right)\left|x\right|^{\delta}\label{eq:3.2}
\end{equation}
and the corresponding Schrödinger equations
\begin{equation}
-\frac{d^{2}\psi_{n}^{\pm}}{dx^{2}}+\left[g^{2}x^{2\left(1+\delta\right)}\pm g\left(1+\delta\right)\left|x\right|^{\delta}\right]\psi_{n}^{\pm}=E_{n}^{\pm}\psi_{n}^{\pm}.\label{eq:3.3}
\end{equation}
Recall that here we are considering $2m=1$. 

Since the ground state wavefunction $\psi_{0}^{-}\left(x\right)$
of $H_{-}\left(=\bar{A}A\right)$ corresponds to a zero energy $E_{0}^{-}=0$,
i. e. $H_{-}\psi_{0}^{-}=0$, this may be found by using the formula
(\ref{eq:2.17c}) along with (\ref{eq:3.1}) or by imposing the condition
$A\psi_{0}^{-}=0$. From (\ref{eq:2.17c}), it follows easily that
\begin{equation}
\psi_{0}^{-}\left(x\right)=N\,\exp\left[-\frac{g}{2+\delta}\left|x\right|^{2+\delta}\right],\label{eq:3.4}
\end{equation}
where $N$ is the normalization constant given by $N=\frac{\left(\frac{2g}{2+\delta}\right)^{1/\left[2(2+\delta)\right]}}{\sqrt{2\Gamma\left[1+1/(2+\delta)\right]}}$.

In order to be able to gain an understanding of the method that will
be employed later on, we compute the superpotential $W\left(x\right)$
assuming that this is unknown through the perturbation $\delta$-expansion.
For this purpose, we first expand the potential $V_{-}\left(x\right)$
and the ``unknown'' superpotential $W(x)$ in powers of $\delta$,
and then substitute these results in the corresponding Riccati equation.
In other words, we are going to solve perturbatively the Riccati equation
in the $\delta$ parameter.

The series expansion of the potential $V_{-}$ is
\begin{align}
V_{-}\left(x\right) & =g^{2}x^{2}-g+\sum_{n=1}^{\infty}\left[\frac{g^{2}x^{2}\ln^{n}\left|x\right|^{2}-g\mbox{ln}^{n}\left|x\right|}{n!}-\frac{g\mbox{ln}^{n-1}\left|x\right|}{\left(n-1\right)!}\right]\delta^{n},\nonumber \\
 & =\left(g^{2}x^{2}-g\right)+\left(g^{2}x^{2}\ln\left|x\right|^{2}-g\ln\left|x\right|-g\right)\delta^{1}+\cdots\label{eq:3.5a}
\end{align}
and assuming as mentioned before that the superpotential $W$ is unknown,
we write it as a power series in $\delta$ with $x$-dependent coefficients
\begin{equation}
W\left(x\right)=\sum_{n=0}^{\infty}\omega_{n}\left(x\right)\,\delta^{n}=\omega_{0}\left(x\right)+\omega_{1}\left(x\right)\,\delta+\omega_{2}\left(x\right)\delta^{2}+\cdots.\label{eq:3.5b}
\end{equation}
As will be seen below, the unknown coefficients $\omega_{n}\left(x\right)$
will be determined by means of the Riccati equation. 

Substituting the expansions (\ref{eq:3.5a}) and (\ref{eq:3.5b})
into the Riccati equation $V_{-}=W^{2}-W'$ and comparing terms with
the same power in $\delta$, one obtains in general an infinity set
of coupled differential equations (except for one independent equation
which results from $\delta=0$ ). Up to order two in $\delta$, this
process leads to
\begin{eqnarray}
\omega_{0}^{2}-\omega_{0}' & = & g^{2}x^{2}-g\label{eq:3.6a}\\
2\omega_{0}\omega_{1}-\omega_{1}' & = & g^{2}x^{2}\mbox{ln}\left|x\right|^{2}-g\,\mbox{ln}\left|x\right|-g\label{eq:3.6b}\\
2\omega_{0}\omega_{2}+\omega_{1}^{2}-\omega_{2}' & = & \frac{1}{2}\left[g^{2}x^{2}\left(\mbox{ln}\left|x\right|^{2}\right)^{2}-g\left(\mbox{ln}\left|x\right|\right)^{2}\right]-g\,\mbox{ln}\left|x\right|.\label{eq:3.6c}
\end{eqnarray}
The method for solving this system of differential equations is sequential,
i. e., one first solves the independent equation (\ref{eq:3.6a})
to find $\omega_{0}$, then with this function at hand solves (\ref{eq:3.6b})
to find $\omega_{1}$, and so on. However, caution is needed here,
for the differential equation (\ref{eq:3.6a}) has a family of solutions:
\begin{equation}
\omega_{0}\left(x\right)=gx-\frac{2\sqrt{g}\,\mbox{e}^{gx^{2}}}{2\sqrt{g}c-i\sqrt{\pi}\,\mbox{erf}\left(i\sqrt{g}x\right)},\label{eq:3.7}
\end{equation}
where $c$ is an arbitrary constant and $\mbox{erf}\left(x\right)=2/\sqrt{\pi}\int_{0}^{x}dy\,\mbox{e}^{-y^{2}}$
is the well-known error function. Therefore to choose the correct
solution $\omega_{0}\left(x\right)$ we must contrast it with the
corresponding one of the unperturbed model which results of taking
$\delta=0$, i. e., the linear harmonic oscillator (LHO). As the ground
state wavefunction $\psi_{0}^{LHO}$ of the harmonic oscillator is
$\psi_{0}^{LHO}\sim\mbox{e}^{-g\, x^{2}/2}$ (in our units), then
the right $\omega_{0}$ solution is $\omega_{0}\left(x\right)=g\, x.$

Inserting the value of $\omega_{0}\left(x\right)$ into (\ref{eq:3.6b})
and using the integration factor $\mbox{e}^{-gx^{2}}$ to simplify
the integration as well as the initial condition $\omega_{1}\left(0\right)=1$,
it is straightforward to show that $\omega_{1}(x)=g\, x\,\mbox{ln}\left|x\right|$.
In a similar manner, using the results for $\omega_{0}\left(x\right)$
and $\omega_{1}\left(x\right)$, one solves (\ref{eq:3.6c}) for $\omega_{2}$,
finding that $\omega_{2}\left(x\right)=(g/2)\, x\,\left(\mbox{ln}\left|x\right|\right)^{2}$.

In a nutshell, we have found perturbatively that the $\delta$-expansion
of the superpotential $W\left(x\right)$ is given by
\begin{equation}
W(x)=\omega_{0}\left(x\right)+\omega_{1}\left(x\right)\delta+\omega_{2}\left(x\right)\delta^{2}+\cdots,\label{eq:3.8a}
\end{equation}
 where, as previously shown,
\begin{equation}
\omega_{0}\left(x\right)=g\, x,\qquad\omega_{1}\left(x\right)=g\, x\,\mbox{ln}\left|x\right|,\qquad\omega_{2}\left(x\right)=(g/2)\, x\,\left(\mbox{ln}\left|x\right|\right)^{2}.\label{eq:3.8b}
\end{equation}
Notice that this expansion coincides (as should be expected) with
that obtained by using the exact form of $W\left(x\right)$ given
in (\ref{eq:3.1}).

We now pass to compute the energy $E_{1}^{-}$ of the first excited
state of $H_{-}$, namely $\psi_{1}^{-}\left(x\right)$, and to this
end we shall use the following trick. Since the ground state wavefunction
$\psi_{0}^{+}\left(x\right)$ of $H_{+}$ is connected by supersymmetry
to $\psi_{1}^{-}\left(x\right)$, $\psi_{1}^{-}\left(x\right)\sim\bar{A}\psi_{0}^{+}\left(x\right)$,
and both have the same energy eigenvalue $E_{1}^{-}=E_{0}^{+}$, we
are going to work with the Hamiltonian $H_{+}$ rather than $H_{-}$,
by refactoring it and then by solving approximately the corresponding
Riccati equation. Let us see below how effectively this trick works.

Considering the ``fermionic'' Hamiltonian $H_{+}=A\bar{A}=-d^{2}/dx+V_{+}$
and following the procedure described in the final part of Sec. \ref{sec: SQM},
we factor $H_{+}$ in the form
\begin{equation}
H_{+}=A\bar{A}=\bar{S}S+\mathcal{E},\label{eq:3.9a}
\end{equation}
where $\mathcal{E}=E_{1}^{-}=E_{0}^{+}$ and
\begin{equation}
S=U\left(x\right)+ip\qquad\qquad\bar{S}=U\left(x\right)-ip,\label{eq:3.9b}
\end{equation}
where $p=-id/dx$. Note that the operators $S$ and $\bar{S}$ play
the same role as $A$ and $\bar{A}$ respectively, whereas $U\left(x\right)$
is a new superpotential to be determined later on and plays the same
role as $W$.

Substituting the definitions of the operators $A$, $S$ and of their
Hermitian adjoints into the second equality of (\ref{eq:3.9a}), we
find a relation between the superpotentials $W$ , $U$ and the energy
$\mathcal{E}$ of the first excited state of $H_{-}$:
\begin{equation}
W^{2}+W'=U^{2}-U'+\mathcal{E}.\label{eq:3.10}
\end{equation}
This relation is a Riccati-like equation and will be solved perturbatively
in the $\delta$ parameter. Analogously to what was done in getting
(\ref{eq:3.8a}), we will assume a power series in the $\delta$ parameter
for all elements involved in (\ref{eq:3.10}). That is,
\begin{eqnarray}
W\left(x\right) & = & \sum_{n=0}^{\infty}\omega_{n}\,\delta^{n}=\omega_{0}\left(x\right)+\omega_{1}\left(x\right)\delta+\omega_{2}\left(x\right)\delta^{2}+\cdots\label{eq:3.11a}\\
U\left(x\right) & = & \sum_{n=0}^{\infty}u_{n}\,\delta^{n}=u_{0}\left(x\right)+u_{1}\left(x\right)\delta+u_{2}\left(x\right)\delta^{2}+\cdots\label{eq:3.11b}\\
\mathcal{E} & = & \sum_{n=0}^{\infty}\varepsilon_{n}\,\delta^{n}=\varepsilon_{0}+\varepsilon_{1}\delta+\varepsilon_{2}\delta^{2}+\cdots.\label{eq:3.11c}
\end{eqnarray}
Inserting these expressions into (\ref{eq:3.10}) and matching the
coefficients of terms with the same power in $\delta$ at both sides
of the equality, we get up to order two in $\delta$ a set of three
differential equations,
\begin{eqnarray}
\omega_{0}^{2}+\omega_{0}' & = & u_{0}^{2}-u_{0}'+\varepsilon_{0}\label{eq:3.12a}\\
2\omega_{0}\omega_{1}+\omega_{1}' & = & 2u_{0}u_{1}-u_{1}'+\varepsilon_{1}\label{eq:3.12b}\\
2\omega_{0}\omega_{2}+\omega_{1}^{2}+\omega_{2}' & = & 2u_{0}u_{2}+u_{1}^{2}-u_{2}'+\varepsilon_{2},\label{eq:3.12c}
\end{eqnarray}
where it should be noted that the functions $u_{i}\left(x\right)$
and the quantities $\varepsilon_{i}$ are unknowns, while the functions
$\omega_{i}$ are given in (\ref{eq:3.8b}). 

Solving (\ref{eq:3.12a}) with $\omega_{0}\left(x\right)=g\, x$,
we find as a possible solution
\begin{equation}
\varepsilon_{0}=2g\qquad\qquad u_{0}\left(x\right)=g\, x.\label{eq:3.13}
\end{equation}
This solution is indeed the right one since it corresponds to the
linear harmonic oscillator which results of taking $\delta=0$. Note
however that there is a family of solutions, for instance, $\varepsilon_{0}=4g$
and $u_{0}=gx-1/x$ constitute also a solution of (\ref{eq:3.12a}).

The next step is to work out $\varepsilon_{1}$ and $\omega_{1}\left(x\right)$
by integrating (\ref{eq:3.12b}). Doing this one arrives at
\begin{equation}
u_{1}\left(x\right)=\mbox{e}^{g\, x^{2}}\int_{0}^{x}e^{-gy^{2}}\left[\varepsilon_{1}-2g^{2}y^{2}\mbox{ln}\left|y\right|-g\left(1+\mbox{ln}\left|y\right|\right)\right]dy,\label{eq:3.14}
\end{equation}
where we have made use of the initial condition $u_{1}\left(0\right)=0$.

Using the boundary condition $u_{1}(x\rightarrow\infty)\rightarrow0$,
which comes from the requirement of finiteness of the wavefunction
at infinity, we obtain directly from (\ref{eq:3.14}) the first contribution
$\varepsilon_{1}$ to the energy $\mathcal{E}$,
\begin{equation}
\varepsilon_{1}=\frac{g\,\left[I\left(0\right)+\,\partial_{\alpha}I\left(0\right)\right]+2g^{2}\partial_{\alpha}I\left(2\right)}{I\left(0\right)}=g\left[\psi\left(3/2\right)-\mbox{ln}g\right],\label{eq:3.15a}
\end{equation}
where $\psi\left(x\right)\doteq d\,\mbox{ln}\Gamma\left(x\right)/dx$,
and
\begin{align}
I\left(\alpha,\, x\right)\doteq\int_{0}^{x}y^{\alpha}\mbox{e}^{-g\, y^{2}}dy & =\frac{1}{2}g^{-\frac{1}{2}\left(1+\alpha\right)}\gamma\left(\frac{1+\alpha}{2},\, g\, x^{2}\right),\label{eq:3.15b}
\end{align}
with $\gamma\left(\alpha,\, x\right)=\int_{0}^{x}\, t^{\alpha-1}\mbox{e}^{-t}dt$
(the incomplete gamma function). Note that for simplicity in (\ref{eq:3.15a})
we have omitted the second argument ($x=\infty$) of the function
$I\left(\alpha,\, x\right)$ .

On the other hand, the function $u_{1}\left(x\right)$ in terms of
$\gamma$ and its first derivative with regard to $\alpha$ is given
by
\begin{equation}
u_{1}\left(x\right)=g\, x\,\mbox{ln}x+\frac{\sqrt{g}}{2}\mbox{e}^{g\, x^{2}}\left[\psi\left(1/2\right)\gamma\left(1/2,\, g\, x^{2}\right)-\partial_{\alpha}\gamma\left(1/2,\, g\, x^{2}\right)\right].\label{eq:3.16}
\end{equation}
With these results at hand the third differential equation (\ref{eq:3.12c})
is tackled. Solving this equation for $u_{2}\left(x\right)$, with
the help of the integration factor $\mbox{e}^{-g\, x^{2}}$and the
initial condition $u_{2}\left(0\right)=0$, we find that
\begin{equation}
u_{2}\left(x\right)=\mbox{e}^{g\, x^{2}}\int_{0}^{x}\, e^{-gy^{2}}\left[\varepsilon_{2}+u_{1}^{2}-2\omega_{0}\omega_{2}-\omega_{1}^{2}-\omega_{2}'\right]dy.\label{eq:3.17}
\end{equation}
Since we are interested only in the value of $\varepsilon_{2}$, at
this point the boundary condition $u_{2}(x\rightarrow\infty)\rightarrow0$
may be used, avoiding so the complex problem of looking for an explicit
solution $u_{2}\left(x\right)$. Hence adopting the notation
\begin{equation}
i_{a}\left(\alpha,\,\beta;\, x\right)\doteq\int_{0}^{x}\,\mbox{e}^{t}\, t^{-a}\gamma\left(\alpha,\, t\right)\gamma\left(\beta,\, t\right),\label{eq:3.18}
\end{equation}
which involves the incomplete gamma function $\gamma$ defined lines
above, and applying the condition $u_{1}(x\rightarrow\infty)\rightarrow0$,
it is easy to verify that
\begin{eqnarray}
\varepsilon_{2} & = & \frac{g}{4\sqrt{\pi}}\biggl\{\psi\left(1/2\right)\left[\partial_{\alpha}i_{1/2}\left(\alpha,\,\alpha\right)-\psi\left(1/2\right)i_{1/2}\left(1/2,\,1/2\right)\right]-\partial_{\alpha\beta}i_{1/2}\left(\alpha,\,\beta\right)\nonumber \\
 &  & +\sqrt{\pi}\left(1+\psi\left(1/2\right)-\mbox{ln}g\right)^{2}+\sqrt{\pi}\biggr\}\biggr|,\label{eq:3.19}
\end{eqnarray}
where $\partial_{\alpha\beta}\doteq\partial_{\alpha}\partial_{\beta}$
and the vertical bar $|$ means evaluation, after performing the respective
differentiations, at $\alpha=\beta=1/2$. As before, for economy in
notation, the third argument ($x=\infty$) of the function $i_{a}\left(\alpha,\,\beta,\, x\right)$
has been dropped.

The majority of the integrals which appear in (\ref{eq:3.17}) were
calculated by reducing them to the master integral ($\ref{eq:3.15b}$)
or to its $\alpha$-derivatives. For instance, an integral like $\int_{0}^{\infty}\mbox{ln}y\,\mbox{e}^{-gy^{2}}$
is simply the derivative of $I(\alpha)$ with respect to $\alpha$,
i. e. $\partial_{\alpha}I\left(\alpha\right)$, evaluated at $\alpha=0$.
By contrast, the remaining integrals in (\ref{eq:3.19}) are very
complex due to the product of two incomplete $\gamma$ functions involved
in the definition of $i_{a}\left(\alpha,\,\beta,\, x\right)$ so that
they have been evaluated numerically. As a result, we have found that
\begin{equation}
\varepsilon_{2}=-0.17638\, g+0.48176\, g\,\mbox{ln}g+0.25\, g\,\left(\mbox{ln}g\right)^{2}.\label{eq:3.20}
\end{equation}
Taking $g=1$ and grouping all the contributions $\varepsilon_{i}$,
the $\delta$ expansion for $\mathcal{E}$ becomes 
\begin{equation}
\mathcal{E}=\varepsilon_{0}+\varepsilon_{1}\delta+\varepsilon_{2}\delta^{2},\label{eq:3.21}
\end{equation}
where $\varepsilon_{0}=2$, $\varepsilon_{1}=0.03649$ and $\varepsilon_{2}=-0.17638$.

If we now focus on the supersymmetric $\varepsilon$-system defined
by $W\left(x\right)=gx^{2}\varepsilon\left(x\right)$ and take so
$\delta=1$ in (\ref{eq:3.21}), we obtain the energy $E_{1}^{-}=1.86011$
for the first excited state of the Hamiltonian $H_{-}$. This result
can be improved by using the $\left[1,1\right]$ Padé approximant
\begin{equation}
\tilde{\mathcal{E}}=\frac{\varepsilon_{0}\varepsilon_{1}+\delta\left(\varepsilon_{1}^{2}-\varepsilon_{0}\varepsilon_{2}\right)}{\varepsilon_{1}-\delta\varepsilon_{2}}.
\end{equation}
For $\delta=1$ an energy of $E_{1}^{-}=2.00626$ for the first excited
state of $H_{-}$ is obtained. This result is closer to its supersymmetric
partner result $E_{0}^{+}=1.94605$ calculated in \cite{Fabricio-Adilson (2012)}
by using a variant of the logarithmic perturbation theory, improved
with the same {[}1,1{]} Padé approximant. Both results must be compared
with the very precise values $E_{1}^{-}=E_{0}^{+}=1.969507538$ obtained
in \cite{Fernandez (2013)} by a variational (Rayleigh-Ritz method)
with the use of a seven parameter trial solution.

\section{Conclusions\label{sec:Conclunsions}}

The purpose of this paper is two-fold. First, we generalize the standard
supersymmetric quantum mechanics by introducing a new class of operators
$A$ and $\bar{A}$, and show that in the linear definition these
operators reduce to the conventional ones proposed in \cite{Bernstein-etal (1984)}.
Higher-order operator formulation is in progress, which for the second
order reduces to those proposed in \cite{Andrianov-Ioffe (1993)}.
Second, we revisit the supersymmetric $\varepsilon$-system introduced
in \cite{Fabricio-Adilson (2012)} and exploit its supersymmetry in
order to determine the first excited state energy of the bosonic Hamiltonian
$H_{-}$. Comparison with the results of \cite{Fabricio-Adilson (2012)}
shows that the logarithmic approximation developed in \cite{Bender (1989),Cooper & Roy (1990)}
does not give better results for the energy levels of the $\varepsilon$-system
than the simpler linear logarithmic approximation used in \cite{Fabricio-Adilson (2012)}. 
\begin{acknowledgments}
This work was partially supported by Conselho Nacional de Desenvolvimento
Científico e Tecnológico (CNPq). The work of E. A. Gallegos has been
supported by CAPES-Brazil.\end{acknowledgments}

\end{document}